\documentclass[]{elsart}
\def\openone{\leavevmode\hbox{\small1\kern-3.8pt\normalsize1}}
\def\<{\langle} \def\>{\rangle}%

\def\Tr{\hbox{Tr}}

\def\nv{{\bf n}}
\def\yv{{\bf y}}
\def\xv{{\bf x}}
\def\Jv{{\bf J}}
\def\kv{{\bf k}}
\def\Yv{{\bf Y}}
\newcommand{\ket}[1]{ | \, #1  \rangle}
\newcommand{\bra}[1]{ \langle #1 \,  |}
\begin{document}
\begin{frontmatter}
\title{A quantum measurement of the spin direction}
\author{G. M. D'Ariano$^{a,b,c}$, P. Lo Presti$^a$, M. F. Sacchi$^a$} 
\address{$^a$ Quantum Optics \& Information Group,
Istituto Nazionale di Fisica della Materia, Unit\`a di Pavia,
Dipartimento di Fisica ``A. Volta'', via Bassi 6, I-27100 Pavia, Italy\\
$^b$ Istituto Nazionale di Fisica Nucleare, Sezione di Pavia \\
$^c$ Department of Electrical and Computer Engineering, Northwestern
University, Evanston, IL  60208} 
\begin{abstract}
We give a first physical model for the quantum measurement of the spin
direction. It is an Arthurs-Kelly model that involves a kind of
magnetic-dipole interaction of the spin with three modes of
radiation. We show that in a limit of infinite squeezing of radiation
the optimal POVM for the measurement of the spin direction is achieved
for spin $\frac{1}{2}$.
\end{abstract}
\end{frontmatter}
\section{Introduction}
In quantum mechanics everybody knows that we can measure the component
of a spin along a given direction---usually a magnetic field. Why is
not possible to measure the ``direction'' of the spin itself, like
what we do in classical mechanics? There must be a way to define the
measurement of the direction of the spin also in quantum mechanics,
otherwise we would face a situation which is inconsistent with what we
normally do in the macroscopic world!  \par The situation is somewhat
similar to what happens for the measurement of position and momentum:
in classical mechanics we can measure both jointly, whereas in quantum
mechanics we learn that we can measure either one or the other. This
is true since only exact (orthogonal) measurements are usually
considered. However, from quantum estimation theory \cite{helstrom} we
also learned that approximate measurements can be
considered as well, and in this way we can define joint measurements
of position and momentum, and, in principle, measurements of the
direction of a spin.  \par The first scheme for a joint measurement of
non-commuting observables was introduced by Arthurs-Kelly
\cite{A-K}. The problem of evaluating the minimum added noise in the
joint measurement of position and momentum---and more generally of a
pair of observables whose commutator is not a c-number---was then
solved by Yuen \cite{Yu}, and a similar approach to the problem has
been followed in Ref. \cite{goodman}.  In the case of two quadratures
of one mode of the electromagnetic field the problem can be phrased in
terms of a coherent Positive Operator-Valued Measure (POVM) whose
Naimark extension introduces an additional mode of the field: this
kind of measurement can be realised by means of a heterodyne detector
\cite{heterodyne}.

The  case  of  a joint measurement of the  angular  momentum  is by
far more difficult than the joint measurement of position and
momentum, and no measurement scheme has appeared in the literature so 
far. Spin  coherent states have been introduced \cite{arecper} that
can be interpreted as continuous overcomplete POVMs \cite{peres}, but
no physical model has been given for it. It has been shown \cite{Appleby}
that such coherent-spin POVM minimizes suitably defined
quantities that represent the precision and disturbance of the
measurement, and that such POVM would also be optimal for estimating
the rotation parameters of a spin \cite{holevo}. Even
discrete-spectrum POVMs for the  joint measurement of the three  
components $J_x$,  $J_y$ and  $J_z$ of the  angular momentum have not
been analyzed yet, and a physical model based on quantum cloning has
been studied in Ref. \cite{clonjoint}.

A model for the realization of the measurement of the spin
direction would be essential in connecting the quantum with the
classical meaning of the angular  momentum itself, whence it is of
great interest to find viable physical schemes to realize such
measurement. The realization of the spin-direction measurement would
also represent a first step toward the achievement of the POVM that is
needed for teleporting an angular momentum \cite{telep}.

In this paper a first physical model for the measurement of the
spin-direction is presented. It is based on a Arthurs-Kelly model that
involves a sort of magnetic-dipole interaction of the spin with three
modes of radiation. As we will see, the model achieves the optimal
POVM for spin $\frac{1}{2}$, in a limit of infinite squeezing of the
radiation modes.

\section{Coherent-spin POVM}
The spin coherent states $|\nv\rangle$, with
$\nv=(\sin\theta\cos\phi,\sin\theta\sin\phi,\cos\theta)$, are
generated by the action of the unitary operator
$V(\nv)=\exp[i\theta(\nv\wedge\kv)\cdot \Jv)]$,
$\kv=(0,0,1)$, on the eigenstate $|-j\rangle$ of $J_3$ with eigenvalue
$-j$. The unitary transformation $V(\nv)$ is a rotation that maps the
direction $\kv$ on the direction $\nv$, where the rotation is performed
continuously along the meridian connecting the two directions. For
this reason one has 
\begin{equation}
V(\nv)J_3 V(\nv)^\dagger= \Jv\cdot \nv
\;,\label{uno}
\end{equation}
so that 
each spin coherent state is the eigenvector of $\Jv\cdot\nv$ with eigenvalue
$-j$. This set of states provides the coherent-spin
POVM \cite{arecper} which is given by 
\begin{equation}
\Pi(d\nv)=\frac{2j+1}{4\pi}|\nv\>\<\nv|d\nv\;. \label{spin_povm}
\end{equation}
This POVM is a continuous and overcomplete POVM that provides an
unbiased estimation of the spin direction. There is no known Naimark
extension of the POVM (\ref{spin_povm})---i. e. an orthogonal
projective realization of the POVM on an enlarged Hilbert space: we
only know that such extension must be infinite-dimensional, since the
POVM spectrum is continuous. We emphasize again that coherent-spin
POVM (\ref{spin_povm}) minimizes suitably defined precision and
disturbance of the measurement \cite{Appleby}, and that it is the
optimal one for estimating the rotation angles of a spin
\cite{holevo}.

\section{The Arthurs-Kelly model}
In the Arthurs-Kelly scheme \cite{A-K} the joint measurement of two
non-commuting observables on a single quantum system was obtained upon
introducing two auxiliary meters. Here we let a spin $j$ interact with
three independent radiation modes $a_1$, $a_2$, $a_3$ for a time
interval $t$, according to the following kind of magnetic-dipole interaction
\begin{equation}
U=\exp\bigg[-it(J_1Y_1+J_2Y_2+J_3Y_3)\bigg]=
\exp\bigg[-it\Jv\cdot\Yv\bigg]\;,
\label{evolutor}
\end{equation}
where $Y_h$ denotes the quadrature $Y_h=\frac{i}2(a_h^\dagger-a_h)$,
and all phases have been included in the definition of 
the annihilation operators $a_h$. We fix the preparation of the radiation
field in the squeezed state
$|\Psi\>=|\psi_\lambda\>_1|\psi_\lambda\>_2|\psi_\lambda\>_3\;,$ with
\begin{equation}
_h\<y|\psi_\lambda\>_h=\left(\frac{2\lambda^2}{\pi}\right)^{\frac14}
e^{-\lambda^2y^2}\;,
\end{equation}
is the wavefunction on the basis of the eigenstates of $Y_h$.  After
the interaction, the measurement of the quadratures
$X_h=\frac{1}2(a_h^\dagger+a_h)\;$ is performed through independent homodyne
detection on each mode.  The outcome of the measurement is a vector
$\xv=x\nv=x(
\sin\theta\cos\phi,\sin\theta\sin\phi,\cos\theta)$,  and the resulting
POVM for the spin system is given by
\begin{eqnarray}
\Pi_\lambda(d\xv)&=&\Tr_{rad}[\;\openone_s\otimes|\Psi\>\<\Psi|\,U^\dagger\,
\openone_s\otimes|\xv\>\<\xv|\,U\;]d\xv=\nonumber\\
&=&\Omega_\lambda^\dagger(\xv)\,\Omega_\lambda(\xv)\,d\xv\;,\label{pil}
\end{eqnarray}
where $\Tr_{rad}$ denotes the trace over radiation modes, $\openone_s$
is the identity for the spin Hilbert space, and
$\Omega_\lambda(\xv)=\<\xv|e^{-it\Jv\cdot\Yv}|\Psi\>$. By inserting
the completeness relation for the operators $Y_h$ in the matrix
element $\<\xv|e^{-it\Jv\cdot\Yv}|\Psi\>$, $\Omega_\lambda(\xv)$ can
be rewritten as follows
\begin{equation}
\Omega_\lambda(\xv)=
\frac1{\pi^{\frac32}}\left(\frac{2\lambda^2}{\pi}\right)^{\frac34}
\int d^3\yv \,e^{2i\xv\cdot\yv}\,
e^{-it\Jv\cdot\yv}\,e^{-\lambda^2|\yv|^2}\;.\label{omega}
\end{equation}
We  are now interested  in  the limit of Eq.  (\ref{omega}) for
infinitely squeezed radiation $\lambda  \rightarrow 0$.

In order to simplify the analysis, we exploit the rotation covariance
of the operator $\Omega_\lambda(\xv)$
\begin{equation}
\Omega_\lambda(x\nv)=
V^\dagger(\nv)\Omega_\lambda(x\kv)V(\nv)\;,\label{omegatrans}
\end{equation}
along with the fact that $\Omega_\lambda(x\nv)$ commutes with
$\Jv\cdot\nv$, namely it is diagonal on the eigenvectors of
$\Jv\cdot\nv$. Therefore, it is enough to evaluate the matrix elements
$\<m|\Omega(x\kv)|m\>$ on eigenstates $|m\>$ of $J_z$. We also know
that such matrix element is real, since $\Omega_\lambda$ is 
self-adjoint. Therefore, we are interested in the following evaluation
\begin{eqnarray}
&&\<m|\Omega_\lambda(x\kv)|m\>=\nonumber\\ &&=
\left(\frac\lambda\pi\right)^{\frac32}\left(\frac2{\pi}\right)^{\frac34}\int
d\nv\int_0^\infty dy\,y^2\,e^{2i\kv\cdot\nv\,x\,y}\, e^{-\lambda^2
y^2}\,
\<m|V(\nv)e^{-it\,J_3\,y}V^\dagger(\nv)|m\>=\nonumber\\
&&=
\left(\frac\lambda\pi\right)^{\frac32}
\left(\frac2{\pi}\right)^{\frac34}\sum_{m'}
\int d\nv\,|\<m|V(\nv)|m'\>|^2
\int_0^\infty dy\,y^2\,e^{-im'ty+2i\kv\cdot\nv \,x\,y}  e^{-\lambda^2 y^2}
\;. \label{step1}
\end{eqnarray}
The function $|\<m|V(\nv)|m'\>|^2$, where $\nv$ points in the
direction $(\theta,\phi)$, does not depend on $\phi$. In fact, if
$\nv'$ is a unit vector pointing in the direction
$(\theta,\phi+\delta)$, one has $V(\nv')=\exp[i\delta
J_3]V(\nv)\exp[-i\delta J_3]$, and thus $|\<m
|V(\nv')|m'\>|^2=|\<m|V(\nv)|m'\>|^2\doteq g^{m'}_m(\cos\theta)$.
Since $\<m|\Omega_\lambda(x\kv)|m\>$ is equal to its real part,
the last integral in Eq. (\ref{step1}) can be replaced with its
real part.  Defining $\eta=\cos\theta$, and integrating on $\phi$,
Eq. (\ref{step1}) becomes
\begin{eqnarray}
&&\<m|\Omega_\lambda(x\kv)|m\>=\nonumber\\
&&=2\pi \left(\frac\lambda\pi\right)^{\frac32}
\left(\frac2{\pi}\right)^{\frac34} \sum_{m'} \int_{-1}^1 
d\eta\,g_m^{m'}(\eta)\,\frac{(-1)}{4x^2}\,\partial_\eta^2\,
\frac12\int_{-\infty}^\infty dy\,
e^{i\,(2\eta x-m't)\,y}\,e^{-\lambda^2\,y}=\nonumber\\
&&=\frac{(-1)}{4x^2}\left(\frac2{\pi}\right)^{\frac34}
\lambda^{1/2}\sum_{m'}\int_{-1}^1 d\eta \,g^m_{m'}(\eta)
\partial^2_\eta \exp\bigg[\frac{-(2\eta x-m't)^2}
{4\lambda^2}\bigg]\;.
\end{eqnarray}
Using the partial integration rule 
$\int g \partial^2 f= g\partial f - f\partial g + \int \partial^2g f$,
and keeping the term $g\partial f $, which is the only one that
survives in the limit $\lambda \rightarrow 0$, one finds
\begin{eqnarray}
\bra{m}\Omega_\lambda(x\kv)\ket{m} 
&\stackrel{\lambda\rightarrow0}{\approx}  & 
  \frac{1}{4x^2}\left(\frac{2}{\pi}\right)^{3/4}{\lambda}^{1/2}
\sum_{m'} |\bra{m}V(\nv)\ket{m'}|^2 
\frac{x}{\lambda^2} \times \nonumber \\
 & \times & (2x\cos\theta_{\nv} - m't)
  \exp\bigg[-\frac{(2x\cos\theta_{\nv}-m't)^2}
   {4\lambda^2}\bigg] \bigg\arrowvert^{\nv=\kv}_{\nv=-\kv}= \nonumber \\
  & = & \frac{1}{2}\left(\frac{2}{\pi}\right)^{3/4}
\frac{2x-mt}{x\lambda^{3/2}}\exp\bigg[\frac{-(2x -mt)^2}{4\lambda^2}\bigg],
\end{eqnarray}
where we used $ |\bra{m}V(\pm \kv)\ket{m'}|^2=\delta_{m,\pm  m'}$.
From this result one has
\begin{eqnarray} 
&&\bra{m}\Omega_\lambda^{\dagger}(x\kv)\Omega_\lambda(x\kv)\ket{m}
  \stackrel{\lambda\rightarrow0}{\approx} \nonumber \\& & 
   \frac{1}{\pi x^2}\left(\frac{1}{\sqrt{2\pi}\lambda}+\
   \frac{\lambda}{4\sqrt{2\pi}}\partial_x^2 \right)
    \exp\left[-\frac{(2x-mt)^2}{2\lambda^2}\right]\;.
\end{eqnarray}
Recalling Eqs. (\ref{pil}), (\ref{omegatrans}), 
and (\ref{uno}), the POVM writes
\begin{equation}
\Pi(d\xv)\approx\frac1\pi \frac{1}{\sqrt{2\pi}\lambda}
    \exp\left[-\frac{(2x\openone-t\Jv\cdot\nv)^2}{2\lambda^2}\right]dx d\nv\;.
\label{popo}\end{equation}
Notice that
\begin{eqnarray}
&&  \bra{m}\Pi(d\xv)\ket{m}
  \rightarrow  \left\{\begin{array}{cc} 
    \frac{1}{\pi }\delta (2x-mt)\;,\;\;& 
     \hbox{if $mt\neq 0$}  \\ \\
     \frac{1}{2\pi }\delta (2x)\;,\;\;& 
     \hbox{if $mt= 0$} \end{array}\right. \;,\label{mOm'0}
\end{eqnarray}
where the two different results are due to the fact that $x\geq0$, so that in
the case of $mt=0$ only a half of the Gaussian must be
considered. Notice also that depending on the chosen sign of $t$, only
a definite sign of $m$ will contribute for $mt\neq 0$. Therefore, in
the limit $\lambda \rightarrow 0$, Eq. (\ref{popo}) becomes
\begin{equation}
\Pi(d\xv)=\frac1{2\pi}\sum_{m,\, mt\ge 0}
\left(1-\frac12\delta_{m,0}\right)\,\delta \left(x-\frac{mt}2\right)
\,|\nv,m\>\<\nv,m|
\, dx \,d\nv
\;.\label{fin}
\end{equation}
Equation (\ref{fin}) shows that only outcome vectors $\xv$ with
$|\xv|=mt/2$ have non vanishing probability, whence we can also easily
express the transformation of the spin state due to the measurement
\begin{eqnarray}
&&\rho\to\frac{I_{dx,d\nv}(\rho)}{\Tr[I_{dx,d\nv}(\rho)]},\nonumber\\
&&I_{dx,d\nv}(\rho)=\Omega(\xv)\,\rho\,\Omega^\dagger(\xv)\,dx\,d\nv
\\
&&=
\frac{1}{2\pi}\sum_{m\,mt\ge 0}\left(1-\frac12\delta_{m,0}\right)\,
\delta\left(x-\frac{mt}2\right)\<\nv,m|\rho|\nv,m\>
|\nv,m\>\<\nv,m|\, dx\, d\nv\;. \nonumber 
\end{eqnarray}
Taking the marginal POVM---i.e. integrating Eq. (\ref{fin}) over
$x$---for $j=1/2$ one has
\begin{equation}
 \Pi'(d\nv)=\frac1{2\pi}|\nv,1/2\>\<\nv,1/2|\,d\nv
\;,\label{12}
\end{equation}
which is the coherent-spin POVM for spin $1/2$. For larger $j\ge
\frac{1}{2} $, instead, we get a ``mixed'' POVM which is not the
optimal coherent-spin one.
\section{Conclusions}
We have presented a first model for a physical realization of the
measurement of the spin direction, corresponding to the coherent-spin
POVM, which also represents the optimal estimation of  spin rotation
angles. In our Arthurs-Kelly model, the spin is coupled with three
radiation modes that are homodyne detected 
after the interaction. In the limit of highly squeezed radiation the
coherent-spin POVM is achieved only for spin $j=1/2$. On the present
model we have seen which difficulties we need to face in a quantum
measurement of the spin direction, and we hope to have opened the
route for a concrete experimental scheme to achieve this new kind of
quantum measurement.
\section{Acknowledgements}
\par This work has been cosponsored by the Istituto Nazionale di
Fisica della Materia through the project PAIS-99 TWIN, and by the
Italian Ministero dell'Universit\`a e della Ricerca Scientifica e
Tecnologica (MURST) under the co-sponsored project 1999 {\em Quantum
Information Transmission And Processing: Quantum Teleportation And
Error Correction}.

\end{document}